\documentclass[conference]{IEEEtran}
\usepackage{subcaption}
\IEEEoverridecommandlockouts
\usepackage{cite}
\usepackage{amsmath,amssymb,amsfonts}
\usepackage{algorithmic}
\usepackage{graphicx}
\usepackage{textcomp}
\usepackage{pgfplots}
\usepackage{xcolor}

\usepackage[bookmarks=false, pdfencoding=auto, draft]{hyperref}

\usepackage[T1]{fontenc}
\def\BibTeX{{\rm B\kern-.05em{\sc i\kern-.025em b}\kern-.08em
    T\kern-.1667em\lower.7ex\hbox{E}\kern-.125emX}}
\pgfplotsset{compat=1.18}

\usepackage{fancyhdr}

\fancypagestyle{firstpage}{

    \fancyhf{} 

    \lhead{\small Published as a conference demo paper at PerCom 2026} 

}

\begin{document}

\title{LiveSense: A Real‑Time Wi‑Fi Sensing Platform for Range–Doppler on COTS Laptop}





\author{\IEEEauthorblockN{Jessica Sanson\IEEEauthorrefmark{1},
Rahul C. Shah\IEEEauthorrefmark{2},
Maximilian Pinaroc\IEEEauthorrefmark{2},
Cagri Tanriover\IEEEauthorrefmark{2} and
Valerio Frascolla\IEEEauthorrefmark{1}}
\IEEEauthorblockA{\IEEEauthorrefmark{1}Intel Deutschland GmbH, Munich, Germany}
\IEEEauthorblockA{\IEEEauthorrefmark{2}Intel Labs, Santa Clara, CA, USA}
\IEEEauthorblockA{Email: \{jessica.sanson, valerio.frascolla, rahul.c.shah, maximilian.c.pinaroc, cagri.tanriover\}@intel.com}}
\maketitle

\thispagestyle{firstpage}
\begin{abstract}
We present \textsc{LiveSense}\,\,—a cross‑platform that transforms a commercial off-the-shelf (COTS) Wi‑Fi Network Interface Card (NIC) on a laptop into a centimetre‑level Range–Doppler sensor while preserving simultaneous communication capability. The laptops are equipped with COTS Intel~AX211 (Wi‑Fi~6E) or Intel~BE201 (Wi‑Fi~7) NICs. \textsc{LiveSense} can (i) Extract fully synchronised channel state information (CSI) at $\geq40$\ Hz, (ii) Perform time–phase alignment and self‑interference cancellation on‑device, and (iii) Provide a real-time stream of range, Doppler, sub‑carrier magnitude/phase and annotated video frames to a Python/Qt Graphical User Interface (GUI). The demo will showcase the ability to detect (i) Distance and radial velocity of attendees within a few metres of the device, (ii) Micro‑motion (respiration), and (iii) Hand‑gesture ranging. To the best of our knowledge, this is the first-ever demo to obtain accurate range information of targets from commercial Wi-Fi, despite the limited $160$ MHz bandwidth. 
\end{abstract}




\maketitle

\section{Introduction}
 While Wi-Fi was originally designed for high-speed data transmission, it also provides an unprecedented opportunity for device-free sensing through Joint/Integrated Sensing and Communication (JSAC/ISAC) \cite{b0}. Several EU-funded projects are tackling the challenging task to provide some solutions to outstanding ISAC open points, such as the MultiX \cite{MultiX} and the 6G-SENSES \cite{6G-SENSES} projects, and part of this work comes out of activities performed under those projects. Active Wi-Fi sensing leverages commercial NICs with dual antennas to realize radar-like capabilities. By transmitting and receiving on separate antennas, 
 not only Doppler estimation for motion detection but also, crucially, range estimation can be enabled. This unlocks a set of new applications, e.g., presence detection, hand-gesture recognition \cite{b10}, device-free activity monitoring in smart environments and extended reality (XR) \cite{b9}.

However, until recently, extracting precise range information from commercial off-the-shel (COTS) Wi-Fi devices was not possible: prior approaches required external Software Defined Radios (SDRs), or could only obtain Doppler velocity. The breakthrough work \cite{SansonRange} demonstrated for the first time a practical method for accurate range and Doppler extraction from COTS Wi-Fi NICs.

\textsc{LiveSense} is the first end-to-end, real-time platform to offer such unique features.
It processes CSI in real time, providing centimetre-level range and Doppler estimation, supporting human presence sensing, range and velocity detection from moving objects as well as breathing detection. It also allows for sub‑decimetre hand‑gesture ranging, despite the limited $160$ MHz Wi‑Fi bandwidth available on low-cost, COTS laptops.

\subsection{Monostatic Sensing and Real-World Generalization}

While earlier Wi‑Fi sensing systems predominantly adopted a bistatic or multistatic operation, requiring environment‑specific calibration, recent works have begun to explore monostatic sensing over Wi‑Fi to overcome this limitation. For instance, {\itshape ISAC‑Fi} \cite{chen2024isacfi} demonstrated a prototype Wi‑Fi device with self‑interference cancellation enabling monostatic sensing under standard communication workloads. Similarly, a recent SDR-based implementation \cite{Kristensen2025monostatic} showed stable, long‑duration human motion sensing (e.g.\ breathing) up to 10m under non‑line‑of‑sight conditions using a single device. Building on these advances, our proposed system further pushes the envelope by operating on unmodified commercial laptops, leveraging 160 MHz bandwidth for high Signal-to-Noise Ratio (SNR), and achieving centimeter‑level range + Doppler estimation; all without requiring per-device or per‑environment calibration.

\begin{table}[b]
\centering
\caption{Comparison of \textsc{LiveSense} vs. Standard Bistatic Wi-Fi Sensing}
\label{tab:comparison}
\resizebox{\columnwidth}{!}{%
\begin{tabular}{|l|l|l|}
\hline
\textbf{Feature} & \textbf{Standard Bistatic Sensing} & \textbf{LiveSense (Proposed)} \\ \hline
\textbf{Topology} & \textbf{2+ Devices} (Tx \& Rx separated) & \textbf{Monostatic} (Single COTS Laptop) \\ \hline
\textbf{Sensing Metric} & Doppler (Velocity) dominant & \textbf{Range} + Doppler (cm-level) \\ \hline
\textbf{Interference} & Susceptible to env. noise & \textbf{$>$25 dB SNR Gain} (via 160MHz) \\ \hline
\textbf{Calibration} & Environment-specific training & \textbf{Calibration-free} (Auto-alignment) \\ \hline
\textbf{Coexistence} & Often drops packets/rates & \textbf{$>$40Hz} stable under Video/Call load \\ \hline
\textbf{Tracking} & Single target focus & Multi-target (Range domain separation) \\ \hline
\textbf{Deployability} & Controlled setup required & In-the-wild deployment (Cafes, Offices) \\ \hline
\end{tabular}%
}
\end{table}
Table \ref{tab:comparison} highlights the advantages of LiveSense over traditional bistatic COTS sensing \cite{RealTimePrevalent}. Unlike bistatic systems that require separated Transmit / Receive (Tx/Rx) pairs and extensive environmental calibration, \textsc{LiveSense} operates monostatically on a single laptop. By leveraging the full 160 MHz bandwidth (512 subcarriers), our system achieves a processing gain of $>25$ dB, significantly enhancing SNR against interference compared to narrowband solutions. Crucially, \textsc{LiveSense}e ensures seamless coexistence: it dedicates specific spatial streams for sensing while maintaining standard Wi-Fi connectivity on others. We validated this under heavy loads (e.g., simultaneous High-Definition (HD) video streaming and video calls), maintaining a stable sensing rate of $\ge 40$ Hz without packet loss, overcoming the resource scheduling challenges typically found in real-time edge sensing \cite{LiST_LowPower}. The real-time phase alignment of the system and static clutter removal eliminate the need for manual calibration, allowing robust operation in 'in-the-wild' scenarios.

We successfully tested our platform on $>10$ distinct device Stock Keeping Units (SKUs) across diverse regulatory domains (EU, USA, Taiwan, etc.) and in uncontrolled environments like busy coffee shops and offices, demonstrating effective multi-target tracking where bistatic approaches often fail.

\section{Processing Pipeline}
\label{sec:background}

\begin{figure}[t]
  \centering
  \includegraphics[width=0.99\linewidth]{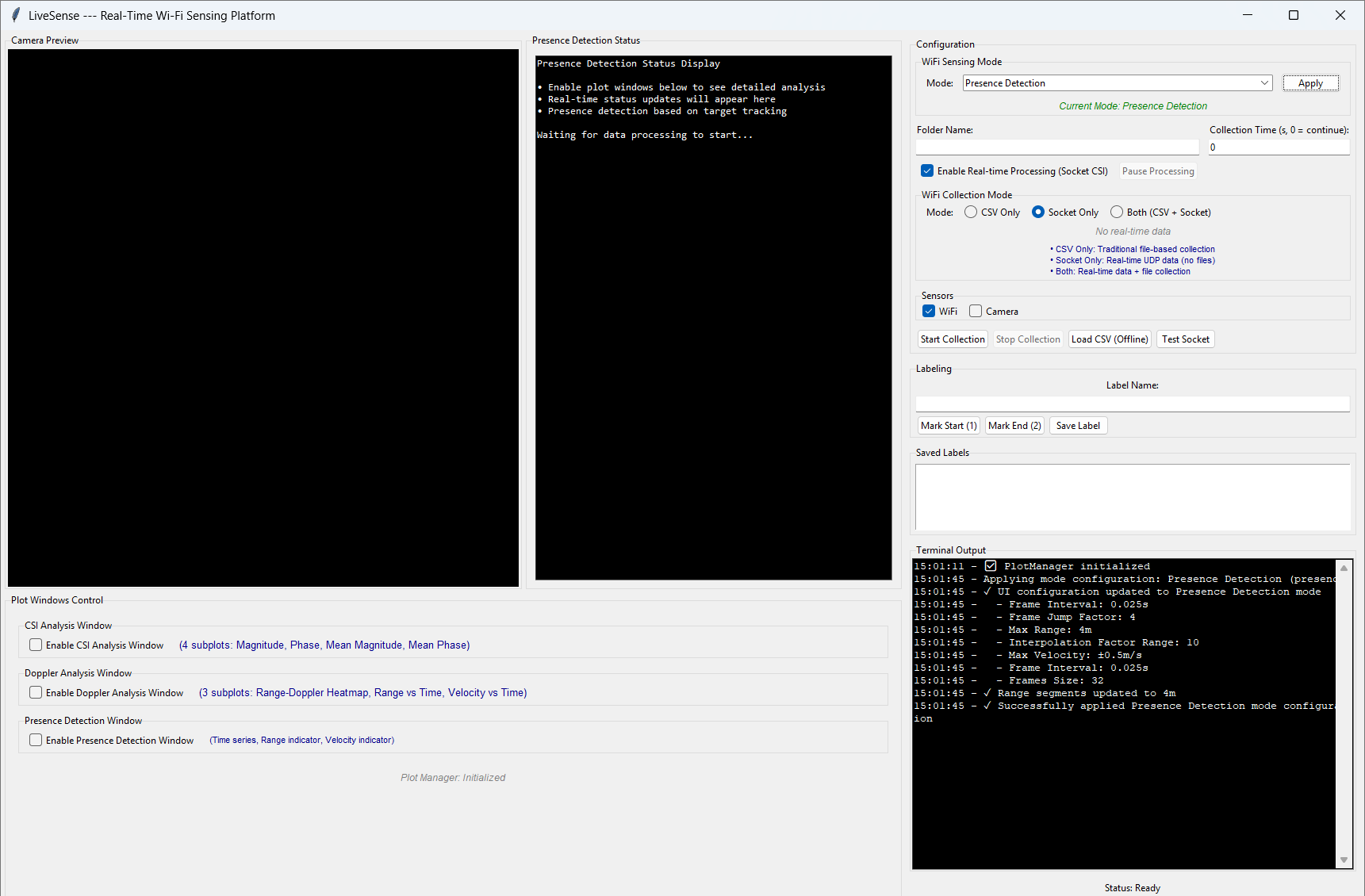}
  \caption{\textsc{LiveSense} graphical user interface}
  \label{fig:arch}
\end{figure}

Real commercial Wi‑Fi hardware (COTS NICs) is not designed for full‑duplex radar‑like operation: hardware asynchronization (e.g., unsynchronized time/phase clocks) and strong Tx–Rx coupling, such as self‑interference (SI), pose serious challenges. To address these, we adapt the signal‑processing techniques recently proposed for  monostatic Wi‑Fi sensing~\cite{SansonRange}.  

Our implementation pipeline performs:

\begin{itemize}
  \item \textbf{CSI Cross‑Correlation} for coarse frame alignment (delay correction), followed by \textbf{sub‑sample interpolation} for fine delay estimation.
  \item \textbf{Per‑Frame Phase Unwrapping and Correction}, using the dominant Tx–Rx coupling (leakage) as a reference — this cancels sampling‑frequency offset (SFO) and phase drift across frames.
  \item \textbf{Adaptive Self‑Interference Cancellation (SIC)}, via a sliding‑window background subtraction (or SI template subtraction) to suppress static clutter (static environment reflections including Tx/Rx coupling) and dynamically adapt to environmental changes. This allows operation without manual recalibration even under moderate environmental variation.
\end{itemize}
\begin{figure}[t]
  \centering
  \includegraphics[width=0.99\linewidth]{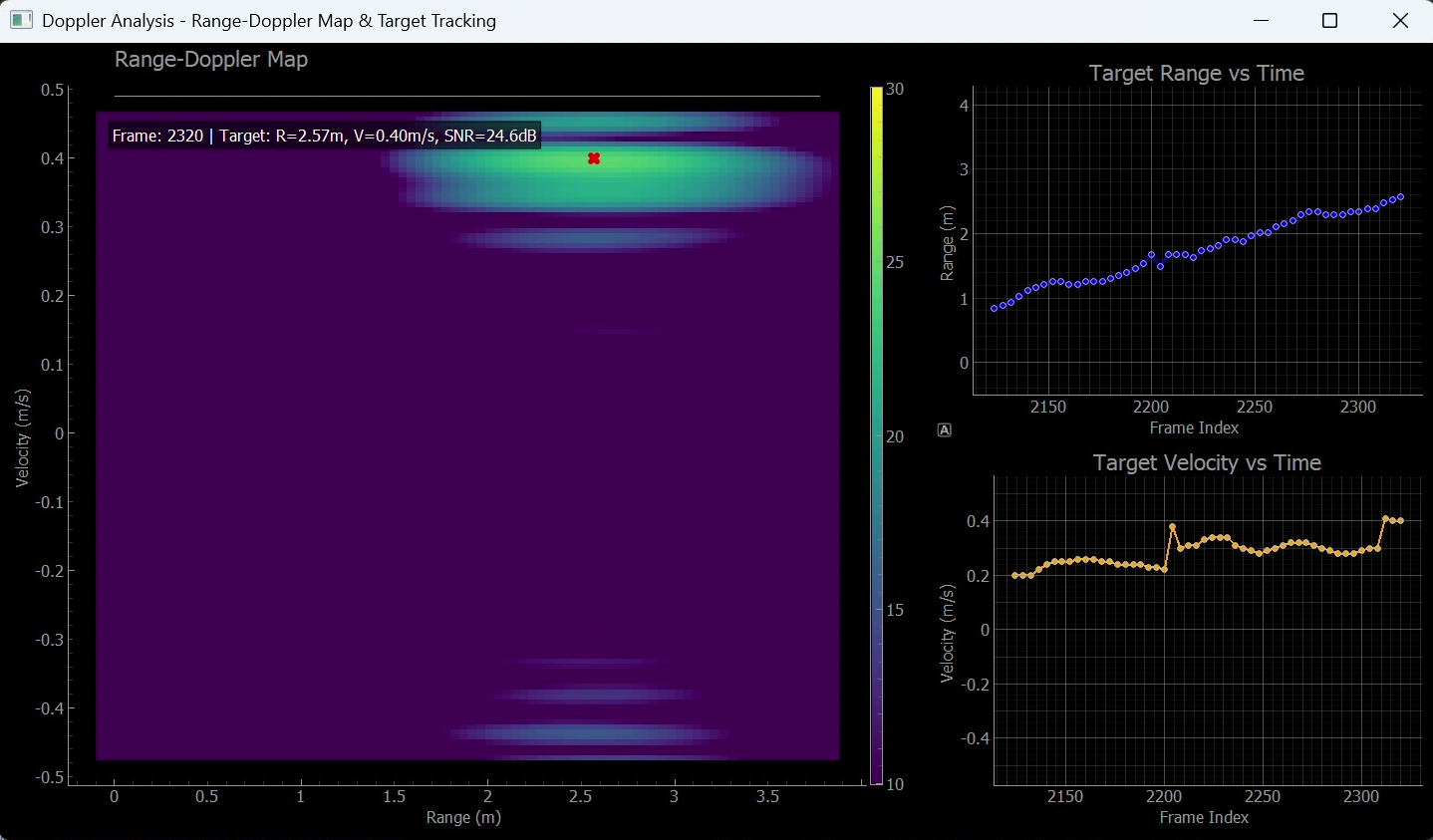}
  \caption{ \textsc{LiveSense in} real-time Range–Doppler heatmap }
  \label{fig:outputs}
\end{figure}

\subsection{Real-Time Sensing Algorithm Pipeline}

Building on this synchronized, cleaned CSI stream, the rest of the pipeline proceeds as follows:

\begin{enumerate}
  \item \textbf{2‑D FFT / DFT}: We apply a Discrete Fourier Transform (DFT) over the \(N\) subcarriers (range domain) and a Fast Fourier Transform (FFT) over the \(M\) temporal frames (Doppler domain).  
  \item \textbf{Detection and Post-Processing}: We estimate SNR and apply a Constant False Alarm Rate (CFAR) detector to identify true reflections from moving targets (people, gestures, micro‑motions such as breathing).  
  \item \textbf{Real-Time Output / Streaming}: The processed range, Doppler and motion estimates are streamed in real time, enabling interactive applications (presence detection, gesture, vital‑sign monitoring, etc.). Our pipeline achieves end-to-end latency of $<1$ $s$ and maintains a stable packet sampling rate of $\ge 40$ $Hz$ under typical communication traffic loads.
\end{enumerate}

To overcome the sparse packet rates typical of standard Wi‑Fi traffic, the system  utilizes \emph{active packet injection} (transmitting dummy/control packets) to sustain a stable sensing sampling rate, ensuring good performance even when the communication load is low.

Figure~\ref{fig:arch} shows the Graphical User Interface (GUI) of \textsc{LiveSense}, where users can configure all sensing and processing parameters and monitor real-time outputs during operation. The platform provides interactive configuration of acquisition and signal-processing parameters, and displays real-time outputs including range–Doppler maps (RDMs), subcarrier statistics, and presence detection. 

All synchronization, range estimation, and pre-processing steps operate on streaming data, while Doppler processing is triggered once $M$ frames are accumulated. For performance stability, we use a buffer size of $4 \times M$ by default, accommodating different application needs and hardware capabilities.

The platform supports three operational modes:
\begin{itemize}
  \item \emph{Gesture mode:} Short range mode that prioritizes high range accuracy via advanced interpolation.
  \item \emph{Presence mode:} Maximizes detection range for occupancy applications.
  \item \emph{Efficiency mode:} Reduces computational overhead for capacity-limited hardware.
\end{itemize}

\textsc{LiveSense} is implemented in Python and can currently handle CSI sampling rates $\geq40$\ Hz on a standard Intel Core~i7‑155U, leaving headroom for GUI rendering and file Input / Output (I/O). Real-time plots include RDMs, historical tracks, phase and magnitude of subcarriers, and presence-detection overlays. Figure~\ref{fig:outputs} presents an example of the real-time output RDM for moving targets from the \textsc{LiveSense} platform.

\begin{figure}[t]
  \centering
  \includegraphics[width=0.99\linewidth]{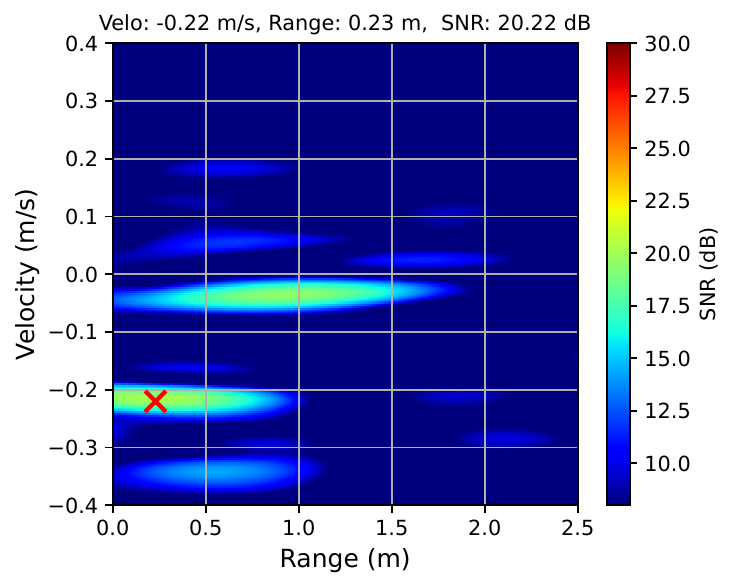}
  \caption{Multi-target estimation in a busy cafe environment. The systemr esolving two distinct targets (0.3 m and 1 m) despite heavy background Wi-Fi traffic, demonstrating robustness against environmental interference.}
  \label{fig:2target}
\end{figure}
The demo uses a Wi-Fi signal with 160 MHz  bandwidth in the 6 GHz band, with 512 subcarriers and a frame interval of 25 ms. Each Doppler batch consists of 32 frames. Under these conditions, the theoretical range resolution is $0.94$ m and the Doppler resolution is $0.03$ m/s, with a maximum unambiguous range of $5$ m and a maximum velocity span of \hbox{$\pm 0.5$ m/s}. In practice, however, LiveSense achieves centimetre-level tracking accuracy through coherent integration, phase compensation, and adaptive interpolation, enabling precise estimation of range and velocity from COTS hardware.


LiveSense was validated across multiple device SKUs and environments. The platform maintains a median range error of $<$ 5 cm for hand gestures ($<$ 0.6 m) and $<$ 20 cm for whole-body motion ($<$ 3 m), with a max range of 4 m. Figure~\ref{fig:2target} demonstrates a crowded café scenario, where the RDM successfully distinguishes a stationary user performing gestures at 0.2 m from a second target walking 1 m behind the device, demonstrating the resilience of the system to background movement and other active Wi-Fi traffic.

\section{Demo Plan}
During the demonstration, attendees are invited to interact directly with the \textsc{LiveSense} platform running on a commercial laptop. The system continuously tracks and displays human presence: as participants approach or move away from the device, a live overlay shows real-time estimates of both their distance and radial velocity. Attendees can also perform hand gestures within a 50\,cm range, allowing the GUI to display the corresponding range estimations and velocity movements.

To showcase sensitivity and robustness of the platform, the demo includes scenarios where participants can stand still in front of the laptop, demonstrating \textsc{LiveSense}'s ability to detect presence and small-scale vital signs, such as breathing. 

\section{Acknowledgments}

This work was partially supported by the European Union’s Horizon Europe SNS JU projects \textsc{6G‑SENSES} (grant 101139282) and \textsc{MultiX} (grant 101192521).


\bibliographystyle{./IEEEtran}

\bibliography{demobib}

\end{document}